\documentclass{article}

\usepackage[numbers]{natbib}
\usepackage[final]{neurips_2019}

\usepackage{placeins}
\usepackage[utf8]{inputenc} 
\usepackage[T1]{fontenc}    
\usepackage[hidelinks]{hyperref}       
\usepackage{url}            
\usepackage{booktabs}       
\usepackage{tabularx}
\usepackage{amsfonts}       
\usepackage{nicefrac}       
\usepackage{microtype}      
\usepackage{graphicx}
\usepackage{xspace}
\usepackage[dvipsnames]{xcolor}
\usepackage{amsmath}
\usepackage{siunitx} 
\usepackage{subcaption}

\newcommand{\worldfloods}{\textit{WorldFloods}\xspace}

\newcommand{\myriad}{Intel\textsuperscript{\textregistered}{}  Movidius\texttrademark{} Myriad\texttrademark{} 2\xspace}
\newcommand{\phisat}{$\Phi$Sat-1\xspace}
\title{Flood Detection On Low Cost Orbital Hardware}

\author{
  Gonzalo Mateo-Garcia\thanks{Equal contribution (listed in alphabetical order)}\\
  University of Valencia\\
 \small{\texttt{gonzalo.mateo-garcia@uv.es}}
  \And
  Silviu Oprea\footnotemark[1]\\
  University of Edinburgh\\
  \small{\texttt{silviu.oprea@ed.ac.uk}}
  \And
  Lewis Smith\footnotemark[1] \\
  University of Oxford \\
  \small{\texttt{lsgs@robots.ox.ac.uk}}
  \AND
  Josh Veitch-Michaelis\footnotemark[1] \\
  Liverpool John Moores University\\
  \small{\texttt{j.l.veitchmichaelis@ljmu.ac.uk}}
  \And
  Guy Schumann\\
  University of Bristol \\
  \small{\texttt{guy.schumann@bristol.ac.uk}}
  \And
  Yarin Gal\\
  University of Oxford \\
  \small{\texttt{yarin@cs.ox.ac.uk}}
  \And
  Atılım Güneş Baydin\\
  University of Oxford\\
  \small{\texttt{gunes@robots.ox.ac.uk}}
  \And
  Dietmar Backes\\
  University of Luxembourg\\
  \small{\texttt{dietmar.backes@uni.lu}}
}

\begin{document}

\maketitle

\begin{abstract}
Satellite imaging is a critical technology for monitoring and responding to natural disasters such as flooding. Despite the capabilities of modern satellites, there is still much to be desired from the perspective of first response organisations like UNICEF. Two main challenges are rapid access to data, and the ability to automatically identify flooded regions in images. We describe a prototypical flood segmentation system, identifying cloud, water and land, that could be deployed on a constellation of small satellites, performing processing on board to reduce downlink bandwidth by 2 orders of magnitude. We target $\Phi$Sat-1, part of the FSSCAT mission, which is planned to be launched by the European Space Agency (ESA) near the start of 2020 as a proof of concept for this new technology.


\end{abstract}

\section{Introduction}

\par Floods are among the most destructive extreme weather events, affecting millions of people each year~\cite{uniceffloods}. Satellite imagery is one of the most important sources of information for disaster response. Optical (visible and infrared) and synthetic aperture radar (SAR) imagery are routinely used to determine flood extent~\cite{serpico2012} and to help direct relief efforts.

\par Many countries do not have direct access to satellite imagery in the event of a disaster. To address this, organisations such as the International Charter ``Space and Major Disasters''\footnote{\url{https://disasterscharter.org}}, initiated by the European Space Agency (ESA), liaise with space agencies and associated  commercial organisations to produce free high resolution maps for users in the field. Despite best efforts it can take many days to provide actionable data, due to satellite tasking and image analysis~\cite{havas2017e2mc}. Commercial organisations are able to provide the highest-frequency (daily) and highest-resolution (sub-metre) images, but these are only freely available for a limited period of time during disaster events. ESA's Copernicus program~\cite{berger2012esa} provides open data globally at \SI{10}{m} resolution, but the optical component, Sentinel 2 (S2,~\cite{drusch2012sentinel}), has a worst-case revisit time of around five days at the equator. This leads to wait periods much longer than two days in areas such as Africa where other alternatives for first response are limited.

\par In this paper we investigate how a constellation of small inexpensive satellites assembled from COTS hardware, also known as CubeSats~\cite{heidt2000cubesat}, could be used for disaster response, using flooding as a case study. 
The main advantage of CubeSats is improved revisit time through larger constellations of satellites. Around 30 CubeSats similar to ESA's upcoming FSSCat mission~\cite{hyperscout2019} could be launched for the cost of a single S2 satellite, reducing the nominal revisit time from 5 days to around 8 hours for the same cost. However, CubeSats can have very limited downlink bandwidth, in the order of 1Mbps. In order to reduce the amount of data being transferred, we suggest performing flood mapping (an image segmentation task) on-board the satellite and only transmitting the final map.

\par We optimised our application for ESA's \phisat, part of the FSSCat mission~\cite{camps2018fsscat}, which is scheduled to be launched in 2020\footnote{FSSCat has been developed by ESA and other partners as a technology demonstrator.}. Among other sensors, FSSCat will carry a HyperScout 2 49-band hyperspectral camera (\SI{80}{m} ground sample distance) which features an \myriad vision processing unit (VPU) as a co-processor for performing on-board computer vision and neural network inference~\cite{hyperscout2019,Esposito19}. Using this capability, a 2-bit flood map (up to 4 classes) would reduce the amount of data being downlinked by a factor of 100 (assuming 49 16-bit channels).

\par The contributions of this paper are as follows:
\begin{enumerate}
    \item We introduce a new dataset, called \worldfloods, containing pairs of Sentinel-2 images and flood extent maps covering 159 global flood events.
    \item We train convolutional neural networks for flood segmentation and compare their performance to standard baseline methods like NDWI.
    \item We design our models to process large volumes of hyperspectral data, yet fit the constraints of hardware deployed on the satellite, and report test results on such hardware.
\end{enumerate}

\section{Flood mapping and related work}


\par Water mapping, of which flood mapping is a special case, is a semantic segmentation task that has been studied for decades. 
A simple approach to water mapping is to compute indices like the Normalised Difference Water Index (NDWI) \cite{McFeeters1996} which exploits the strong absorption of light by water bodies in the green and infrared part of the electromagnetic spectrum. However, this method can perform poorly because the spectral profile of flood water varies widely due to the presence of debris, pollutants and suspended sediments~\cite{MEMON201599}. 

\par More sophisticated segmentation techniques include rule-based classifiers~\cite{MEMON201599} which use fixed or tuned threshold on indices; classical supervised machine learning~\cite{serpico2012}; and recently deep learning~\cite{Rudner2018, isikdogan2017surface}. State-of-the-art results in image segmentation are now routinely achieved using fully convolutional neural networks (FCNNs)~\cite{chen2018encoder}. Most segmentation networks can be described as encoder-decoder architectures and include the popular U-Net~\cite{ronneberger2015u} among many others~\cite{garcia2017review}.


\section{World Floods dataset}

\par The development and evaluation of flooding response systems has been constrained so far by use of datasets of limited geographical scope, with studies often only considering a single flood event \cite{schumannneed}. It is unclear whether such models would accurately generalise to the rest of the world due to variations in topography and landcover. To address this we collated a new global dataset called \worldfloods, which we believe is the largest of its kind.

\par
\begin{figure}[t]
    \centering
    \includegraphics[width=\textwidth]{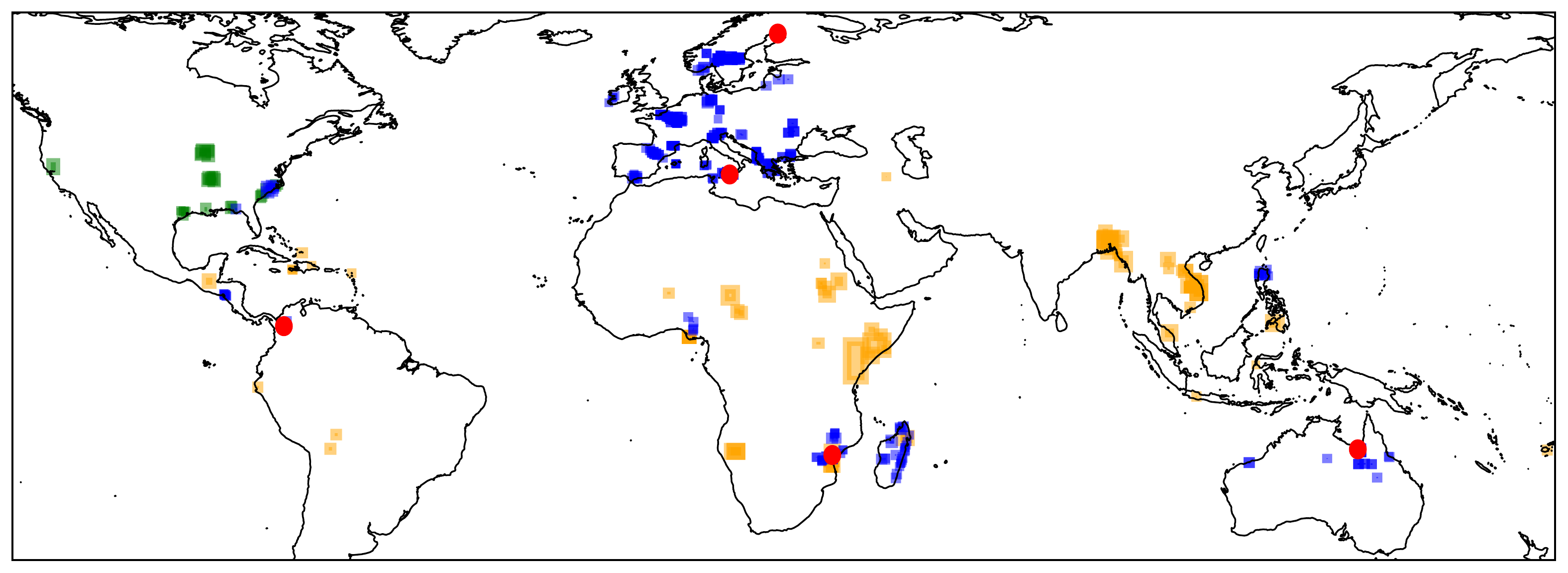}
    \caption{Locations of flood events contained in \worldfloods. Blue, orange and green areas denote Copernicus EMS, UNOSAT and GLOFIMR data, respectively. Red points denote test regions.}
    \label{fig:worldfloods_map}
    \vspace{-4mm}
\end{figure}

 \worldfloods contains 564 flood extent maps created either manually, or semi-automatically, where a human validated machine-generated maps. The dataset covers 159 floods that occurred between November 2015 and March 2019. We sourced all maps from three organisations: the Copernicus Emergency Management Service (Copernicus EMS) \cite{copernicusEMS}, the flood portal of UNOSAT \cite{unosat}, and the Global Flood Inundation Map Repository (GLOFIMR) \cite{glofimr}. The geographical distribution of flood maps is shown in Figure~\ref{fig:worldfloods_map}.

\par For each flood event we provide the raw 13-band S2 image closest in time after the event, and a rasterised segmentation ground truth (cloud, water and land) at \SI{10}{m} resolution. We generated cloud masks using \texttt{s2cloudless}.\footnote{\url{https://github.com/sentinel-hub/sentinel2-cloud-detector}} We manually validated the data to account for gross errors such as missing water bodies or invalid intensities. Sources of error include label noise and temporal misalignment, e.g. the closest S2 image may have been acquired 5--6 days after the map was produced. 
While the labels in the training set may be noisy, we wanted to ensure that the test set provides a fairly clean measurement of the true performance of our system. In this direction, we manually selected test images from flood extent maps that were derived from S2 images and had no misalignment. To avoid data leakage, we did not include test set countries in the training set. Additionally, our test set does not contain any flood maps from events in the training and validation sets. Table \ref{tab:tablestats} shows the train/validation/test statistics.

\begin{table}[ht]
    \centering
    \vspace{-2mm}
    \footnotesize
    \caption{General statistics of the training, validation and test splits of the \worldfloods~dataset.}
    \label{tab:tablestats}
    \begin{tabularx}{\textwidth}{p{12mm}p{10mm}p{10mm}XXXXX}
    \toprule
     Dataset & Flood events & Flood maps & 256x256 patches & Water/flood pixels (\%) & Land pixels (\%) & Cloud pixels (\%) & Invalid pixels (\%) \\
     \midrule
     Training & 60 & 378 & 76,500 & 2.58 & 37.19 & 56.35 & 3.87 \\
     Validation & 6 & 7 & 1,552 & 8.48 & 76.17 & 13.27 & 2.07 \\
     Test & 5 & 10 & 2,029 & 21.14 & 59.57 & 15.98 & 3.31 \\
     \bottomrule
    \end{tabularx}
\end{table}



\section{Experiments}

In order to demonstrate that a FCNN-based flood detection model can segment floods accurately and could be deployed on \phisat, we first train FCNN models on \worldfloods at its original resolution (\SI{10}{m}). We then train models on \textit{degraded} imagery, mimicking the resolution of HyperScout-2 (80 m). We also verify our trained (degraded) models can be run on a \myriad and measure the processing speed.

We focus on the segmentation accuracy of the water/flood class by measuring precision, recall and the Intersection over Union (IoU). Since missing flooded areas (false negatives) is more problematic than over-predicting floods (false positives), high recall is preferred to high precision. Provided the recall was over 95\%, we found that the IoU was a good compromise. 

As baselines, we use NDWI (S2 band 2 and 8) and a linear model (all S2 bands) trained on \worldfloods. A range of NDWI thresholds have been suggested~\cite{McFeeters1996,MEMON201599,rs5073544}. In order to set a stronger baseline, we also compute results where the threshold is tuned for each image, representing the absolute best case performance for the NDWI. We compare our baselines to two FCNNs: a simple CNN (SCNN) comprising four convolutional layers (0.26M parameters) and a U-Net (7.8M parameters,~\cite{ronneberger2015u}). 


Models were trained from scratch for 40 epochs using all 13 S2 bands with input patches of size 256x256 for \SI{10}{m} data or 64x64 for \SI{80}{m} data (2.5 km x 2.5 km). In order to achieve models with high recall we used a cross-entropy loss function that weights each class by the inverse of the observed frequency in Table~\ref{tab:tablestats}, combined with a Dice loss ~\cite{sudre2017generalised}. Augmentation was applied during training including flips and rotations, per-channel jitter, Poisson noise and brightness/contrast adjustments.

\section{Results}

\begin{figure}
\centering
\begin{minipage}{.49\textwidth}
  \centering
  \includegraphics[width=\linewidth]{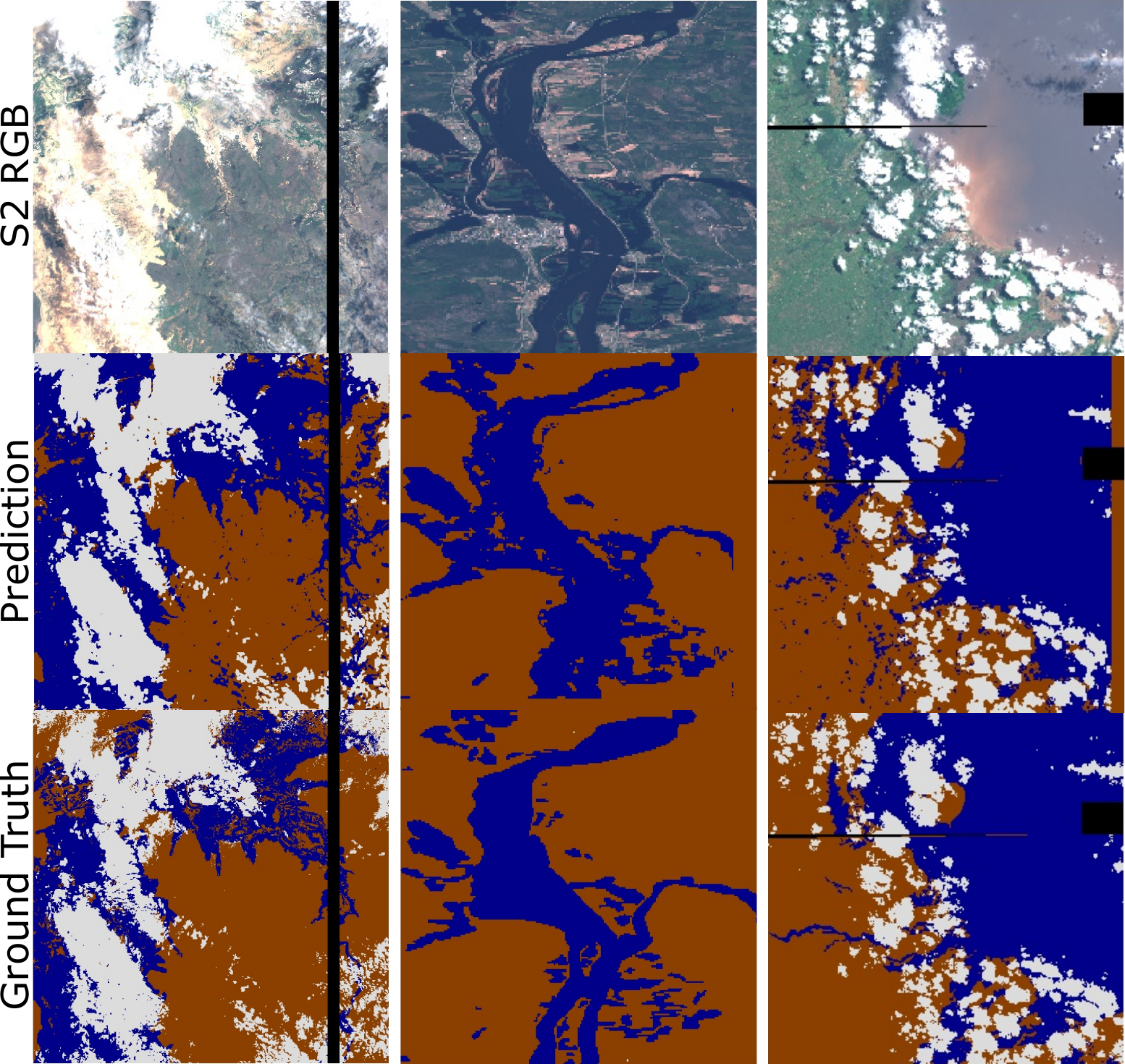}\\
    \captionof{figure}{Segmentation results of degraded models run on \myriad. Black regions are mosaicing artifacts.}
    \label{fig:imgmyriad}
\end{minipage}%
\hspace{.05\textwidth}
\begin{minipage}{.44\textwidth}
  \centering
    \includegraphics[width=0.8\linewidth]{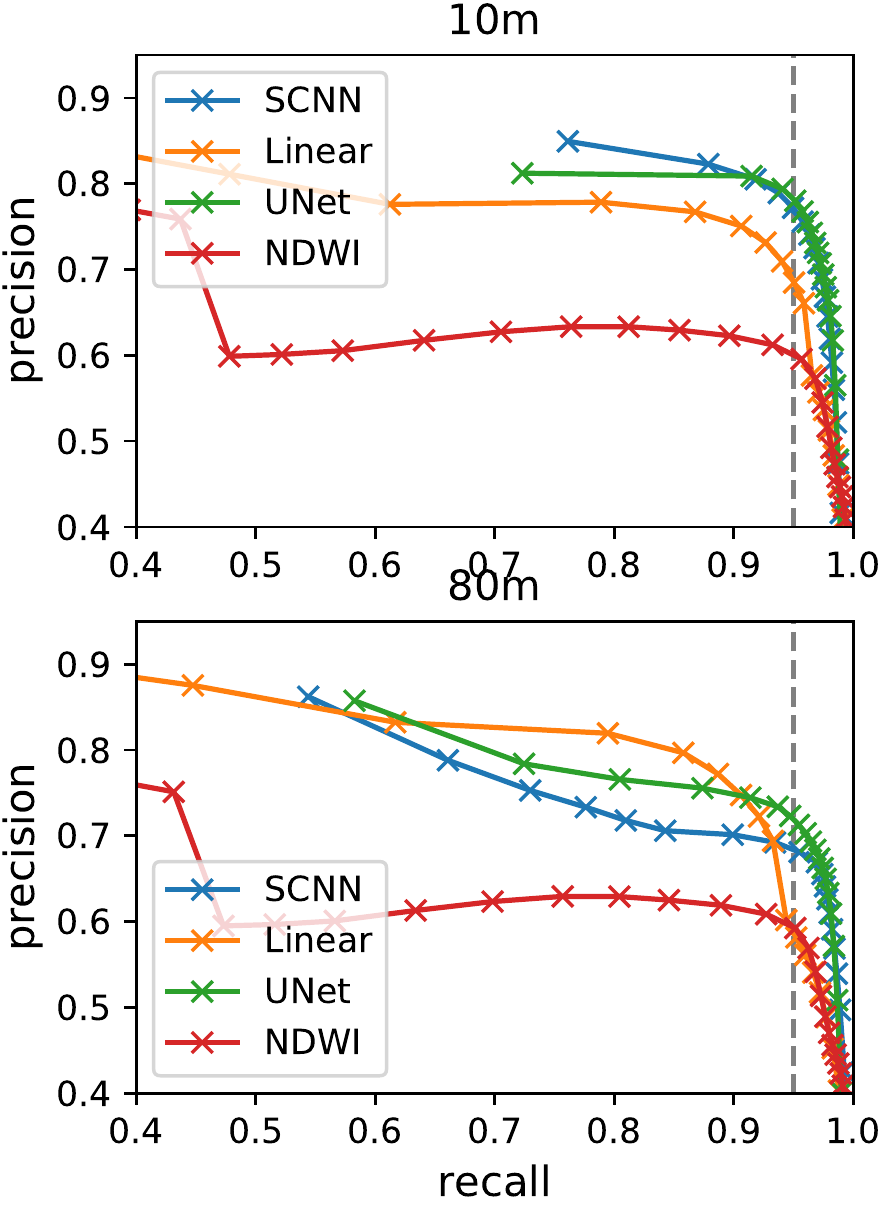}
    \captionof{figure}{Precision-Recall curves of models trained on the S2 resolution (\SI{10}{m}) and HyperScout-2 resolution (\SI{80}{m}). In gray, 95\% recall threshold.}
    \label{fig:prcurve}
\end{minipage}
    \captionof{table}{IoU and recall results for models trained on \worldfloods.}
\label{tab:results}
\footnotesize
    \begin{tabularx}{\textwidth}{p{22mm}p{22mm}p{22mm}XXX}
    \toprule
     {} & NDWI (thresh 0)& NDWI (tuned) & Linear & SCNN  & U-Net \\
     \midrule
     IOU@\SI{10}{m} & 38.37 & 58.63 &  63.34 & 67.54 & \textbf{69.25}  \\
     IOU@\SI{80}{m}  & 37.74 & 58.09 & 57.25 & 65.42 & \textbf{67.51} \\
      \midrule
     Recall@\SI{10}{m} & 43.7 & 93.26 & 96.24 & \textbf{97.43} & 97.34 \\
     Recall@\SI{80}{m}  & 43.13 & 92.72 & 95.24 & \textbf{97.03} & 96.56\\
     \bottomrule
    \end{tabularx}
    \vspace{-5mm}
\end{figure}

Table~\ref{tab:results} shows the IoU and recall for the different models and baselines. Our three models (Linear, SCNN and UNet) all have a recall above 95\% whereas the tuned NDWI has a recall of 93.5\%; NDWI without tuning generalises poorly, we suspect due to muddy water. FCNN models performed best although there was only a small increase in performance between SCNN and U-Net, despite U-Net having 30x more parameters. The drop in performance from \SI{10}{m} to \SI{80}{m} is around 2 points for FCNN models which is acceptable taking into account that the spatial resolution is 8 times worse. Figure~\ref{fig:prcurve} shows the precision and recall for different thresholds on the water/flood class; again, our trained models beat NDWI and larger models tend to perform better.

The SCNN model was selected to be tested on the Myriad 2 chip due to its lower computational footprint compared to UNet (1FLOPS vs 2.68FLOPS for a 64x64x13 input). Figure~\ref{fig:imgmyriad} shows the images segmented using the Myriad 2. In general, the model over-predicts water content. False positives are mostly clustered in the surroundings of water bodies and in cloud shadows. This model segments a 12Mpx image approximately the size acquired by HyperScout-2 in less than one minute.

\section{Conclusion}
We have demonstrated that accurate flood segmentation is feasible to perform using low resolution images in orbit using available hardware. Our models outperform standard baselines and are favourably comparable to human annotation, while being efficiently computable on-board with current hardware. We are currently performing a more rigorous hyper-parameter search over a larger number of models and we hope to release both pipeline code and the \worldfloods~dataset shortly, which we hope will serve as a useful tool to foster research in disaster response.

\subsubsection*{Acknowledgments}
This research was conducted at the Frontier Development Lab (FDL), Europe. The authors gratefully acknowledge support from the European Space Agency, Google Inc., Kellogg College, University of Oxford and other organisations and mentors who supported FDL Europe 2019. 
Gonzalo Mateo-Garcia has been partially supported by the Spanish Ministry of Science, Innovation and Universities (MINECO, TEC2016-77741-R, ERDF) and the European Social Fund.
\FloatBarrier
\bibliographystyle{unsrt}

\bibliography{main}
\end{document}